\documentclass[a4paper,fleqn,usenatbib]{mnras}

\title[tSZ: Forecasts and Cosmology-Dependent Selection]{Forecasts for Next Generation tSZ Surveys: the Impact of a Cosmology-Dependent Selection Function}

\author[Gupta, Porciani, Basu] {Nikhel Gupta\thanks{nikhel.gupta@unimelb.edu.au}$^{1}$,
Cristiano Porciani $^{2}$,
Kaustuv Basu $^{2}$\\
$^{1}$ School of Physics, University of Melbourne, Parkville, VIC 3010, Australia \\
$^{2}$ Argelander-Institut f\"ur Astronomie, Auf dem H\"ugel 71,D-53121 Bonn, Germany \\
}

\usepackage{graphicx}	
\usepackage{amsmath,amsfonts,amssymb}
\usepackage{makeidx}
\usepackage{epstopdf} 
\usepackage{booktabs,caption,fixltx2e}
\usepackage[flushleft]{threeparttable}
\usepackage{hyperref}
\usepackage{multirow}
\usepackage{rotating}
\usepackage{color}
\usepackage[T1]{fontenc}
\usepackage{ae,aecompl}
\usepackage{soul}
\usepackage{mathrsfs}

\definecolor{ored}{rgb}{1.00,0.27,0.00}

\definecolor{Gray}{gray}{0.5}
\definecolor{LightCyan}{rgb}{0.88,1,1}

\def \BE{\begin{equation}}
\def \EE{\end{equation}}	
\def \BC{\begin{center}}
\def \EC{\end{center}}
\def \BEA{\begin{eqnarray}}
\def \EEA{\end{eqnarray}}

\def \SIGMA8{\sigma_{8}}
\def \OM{\Omega_{\rm M}}

\def \rd{{\rm d}}

\usepackage[utf8x]{inputenc}
\usepackage{cellspace}
\begin{document}\sloppy\sloppypar\raggedbottom\frenchspacing

\maketitle

\begin{abstract}
The thermal Sunyaev-Zel'dovich (tSZ) effect is one of the primary tools for finding and characterizing galaxy clusters. 
Several ground-based experiments are either underway or are being planned for mapping wide areas of the sky at $\sim 150$ GHz with large-aperture telescopes. 
We present cosmological forecasts for a  `straw man' tSZ survey that will 
observe a sky area between $200$ and $10^4$ deg$^2$ to
an rms noise level between 2.8 and 20.2 $\mu$K-arcmin.
The probes we consider are the cluster number counts (as a function of
the integrated Compton-$Y$ parameter and redshift) and their angular clustering (as a function of redshift).
At fixed observing time,
we find that wider surveys constrain cosmology slightly better than deeper ones due to their increased ability to detect rare high-mass clusters.
In all cases, we notice that adding the clustering information does not
practically improve the constraints derived from the number counts.
We compare forecasts obtained by sampling the posterior distribution
with the Markov-chain-Monte-Carlo method against those derived using the Fisher-matrix formalism. 
We find that the latter produces slightly optimistic constraints where errors are underestimated at the 10 per cent level. 
Most importantly,
we use an analytic method to estimate the selection function of the survey and account for its response to variations of the cosmological parameters in the likelihood function.
Our analysis demonstrates that neglecting this effect (as routinely done in the literature)
yields artificially tighter constraints by a factor of 2.2 and 1.7 for $\sigma_8$ and $\Omega_\mathrm{M}$, respectively.
\end{abstract}

\begin{keywords}
surveys; methods: statistical; galaxies: clusters: general; cosmology: cosmological parameters
\end{keywords}

\section{Introduction} 

Galaxy clusters play a key role in testing cosmological models.  Measurements of their number counts are helpful to precisely determine the mass and energy contents of the Universe as well as to place tight constraints on the normalization of the linear matter power spectrum and the equation of state of dark energy \citep[e.g.][]{mantz08, vikhlinin09, vanderlinde10, mantz14, planck15, dehaan16, bocquet19, constanzi19}. 
The thermal Sunyaev-Zel'dovich \citep[tSZ,][]{sunyaev70, sunyaev72} effect
is one of the primary tools for detecting and characterizing galaxy clusters today. 
With this name we indicate
a spectral distortion of the cosmic microwave background (CMB) radiation due to the inverse Compton scattering of a small fraction of the photons off hot thermal electrons in the intracluster medium.
The tSZ effect changes the brightness of the CMB along the line of sight to a cluster in a frequency-dependent way. 
Below 218 GHz, an arcminute-scale temperature decrement is produced while an excess is seen at higher frequencies with a peak around 370 GHz. 
Typical temperature variations range between a few tens to $\sim 100$ $\mu$K. 
The tSZ signal (integrated over the area) is proportional to the total thermal energy of galaxy clusters. Moreover, its surface brightness remains unchanged with redshift. These two unique properties lead to an almost redshift-independent mass selection for the tSZ cluster surveys provided that the angular sizes of the telescope beam and clusters on the sky are comparable.

Since the first blind tSZ detections of galaxy clusters \citep{staniszewski09}, 
the field has exploded in the last one decade thanks to several successful ground-based \citep[e.g. the Atacama Cosmology Telescope (ACT) and the South Pole Telescope (SPT), see][]{fowler07, carlstrom11} and also space \citep[e.g. Planck, see][]{planck06} survey experiments. Currently well over 1000 galaxy clusters have been confirmed from their tSZ signal \citep[e.g.][]{planck13-29, hasselfield13, bleem15, hilton18, huang19, bleem19}. Particularly, for the ground-based experiments, the progress is being driven by the development of large-format bolometer cameras with $10^3$-$10^5$ pixels.  
With such wide-field cameras, it is possible to reduce the brightness sensitivity below 10 $\mu$K-arcmin\footnote{The rms (CMB or thermodynamic) temperature variation within a 1 arcmin$^2$ pixel assuming white noise, as commonly used to characterize the sensitivity of CMB experiments.},
which enables cluster detections down to a mass limit of roughly $2\times 10^{14}$ M$_{\odot}$ at any redshift (roughly 1 cluster per deg$^2$). Currently two such experiments are underway: from the Atacama desert \citep[AdvancedACT, see][]{henderson16} and from the South Pole \citep[SPT-3G, see][]{benson14}. Driven by the quest for the CMB primordial B-mode, more experiments are coming online: e.g. the Simons Array \citep{simons-array16}, CCAT-prime \citep[CCAT-p;][]{stacey18,aravena19}, and the Simons Observatory \citep{simons19}. There is also a future plan to merge several such experiments to develop a fourth-generation CMB survey \citep[CMB-S4,][]{cmbs4-19} that will detect a vast number of tSZ clusters. The term `tSZ survey' used in this work merely refers to these multi-frequency CMB experiments that enable blind detections of galaxy clusters via the tSZ effect.

The goal of this paper is to present a simple, yet accurate, scheme for forecasting the constraints on cosmological parameters that will be set from future observations with similar ground-based tSZ surveys. 
In doing this, we show that assuming a fixed limiting mass for the tSZ detections \citep[i.e. that does not vary with the cosmological parameters, as routinely done in previous studies, e.g.][]{holder01b, pierpaoli12, khedekar13} leads to unrealistically optimistic constraints.
We base our study on the published version of a 1000 $\deg^2$ survey planned for the CCAT-p telescope \citep{mittal18, erler18} 
but our results can be interpreted as representative of any similar $\sim 1$-arcmin resolution, ground-based tSZ survey. 
We consider two statistics for the clusters --namely, their number counts
(as a function of the strength of the tSZ signal and redshift) 
and angular 2-point clustering (as a function of redshift)-- but we do not explore combinations with additional datasets (e.g. cross-correlations with other surveys, lensing-based mass estimates, etc.).

The layout of the paper is as follows. In Section \ref{sec:methods}, we introduce the forecasting techniques we use and present a simple analytical method to estimate the survey selection function. 
We also introduce the specifics of our fiducial survey based on a bolometer array for the CCAT-p telescope and then outline wider and/or deeper options. 
In Section \ref{sec:results}, we present our forecasts and compare results obtained using different assumptions and survey areas.
In particular, we discuss the necessity of varying the selection function in accordance with the cosmological parameters. Moreover, we cross check
forecasts obtained using the Fisher information matrix against a full Markov chain Monte Carlo (MCMC) simulation.
In Section \ref{sec:discussion}, we critically discuss our main findings and present our concluding remarks.

Throughout this work, we consider a background flat Friedmann-Lema\^itre-Robertson-Walker cosmological model where the dark-energy and matter density parameters satisfy $\Omega_{\rm DE}+\Omega_{\rm M}=1$. 
Therefore, $\Omega_{\rm M}$, the Hubble constant $H_0=70\,h_{70}$ km s$^{-1}$ Mpc$^{-1}$, and the equation-of-state parameter of dark energy, $w$ (assumed to be redshift independent), fully determine the background. The power spectrum of linear scalar fluctuations, instead, is specified in terms of 
the baryon density parameter $\Omega_{\rm b}$, the scalar spectral index 
$n_s$, and the linear mass variance 
within $11.4\,h_{70}^{-1}$ Mpc spheres 
at redshift zero, $\sigma_8$. We use
the symbols $c, h$ and $k_\mathrm{B}$ to denote the speed of light in vacuum, Planck's constant and Boltzmann's constant, respectively.

\section{Method}
\label{sec:methods}

\subsection{Cluster definition}

The boundary of a cluster is conventionally
defined in terms of the radius $R_\Delta$
within which the (spherically averaged) matter density exceeds
the critical density of the Universe ($\rho_{\rm crit}$) by a factor of $\Delta$. We adopt $\Delta=500$ and consistently define the cluster mass as
\BE
M_{500} = \frac{4}{3} \pi \,500\, \rho_{\rm crit}(z) R_{500}^3\;,
\label{EQ:MDELTA}
\EE
where $z$ denotes the cluster redshift.

\begin{figure*}
\centering
\includegraphics[width=8.5cm]{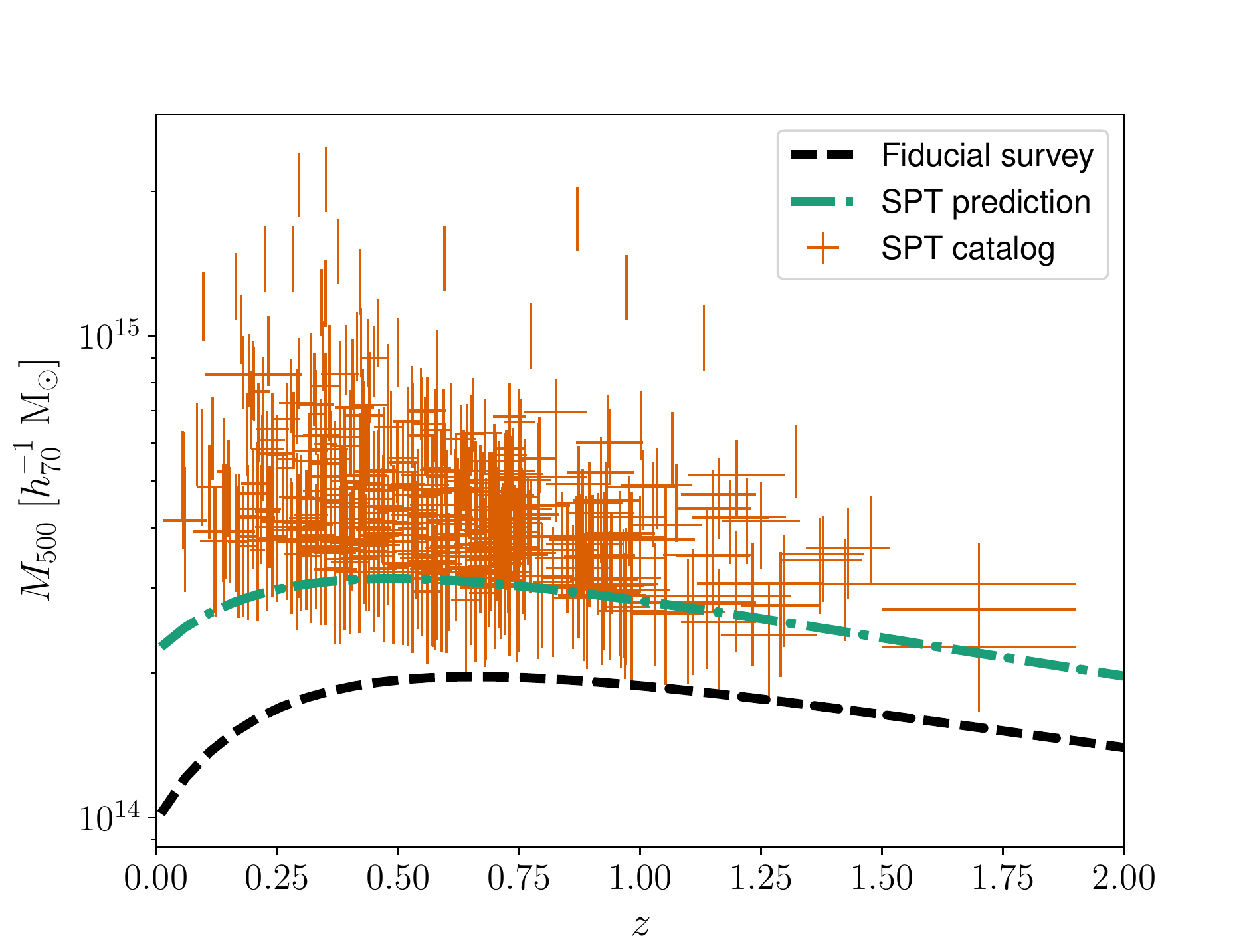}
\includegraphics[width=8.5cm]{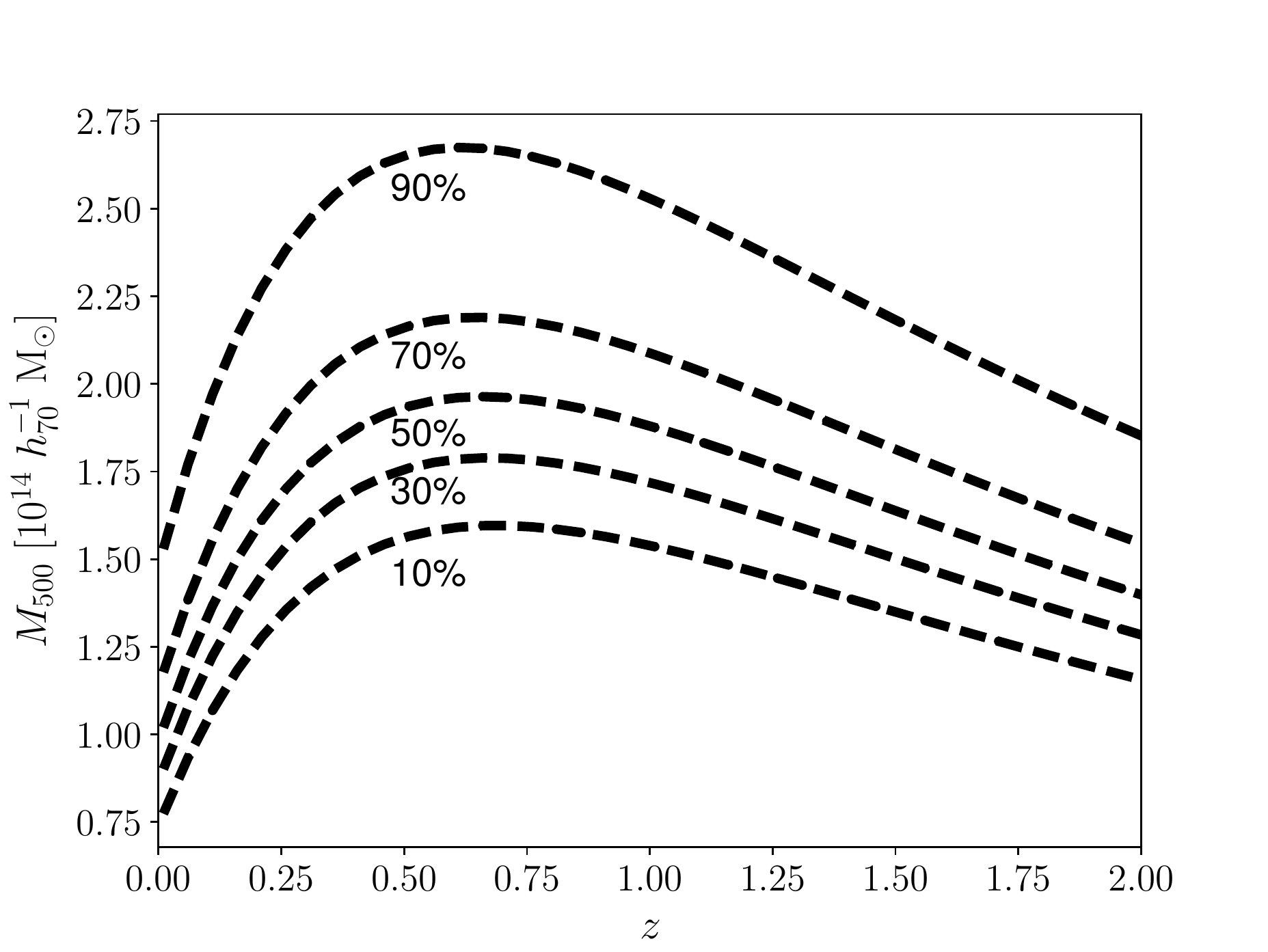}
\caption{Left: The dot-dashed curve shows the redshift-dependent limiting mass obtained from equation (\ref{EQ:COMPLETENESS}) for the SPT survey assuming 50 per cent completeness. The beam noise and size for SPT are 18 $\mu$K and $1.1^{\prime}$, respectively. The data points correspond to the 677 galaxy clusters with $S/N\geq4.5$ in the SPT catalog \citep{bleem15}. The dashed curve is the limiting mass at 50 per cent completeness for our fiducial survey described in Table~\ref{TAB:PARAM}.
Right: Limiting mass estimates for our fiducial survey at 10, 30, 50, 70 and 90 per cent completeness.}
\label{LIMITING_MASS}
\end{figure*}

\subsection{Estimating the survey selection function}
\label{sec:limmass}
Any survey only detects a subsample of the existing clusters due to observational limitations. The sensitivity limit of the survey translates into a limiting mass, above which clusters are detected with a certain signal-to-noise ratio. 
In this section, we present a simple analytical approach to estimate the survey limiting mass.
The change in CMB temperature due to the tSZ effect in a cluster is
\BE
\frac{\Delta T_{\mathrm{SZ}}(\theta)}{T_{\mathrm{CMB}}}=g_{\nu}\frac{\sigma_{T}}{m_{\rm e}c^{2}}\int P_{\rm e}\left(\sqrt{\ell^{2}+\theta^{2}D_{\rm A}^{2}}\right){\mathrm d}\ell,
\label{EQ:TEMP-DECR}
\EE
where $\theta$ is the angular distance from the center of the galaxy cluster, $\ell$ is the radial coordinate from the cluster center along the line of sight, $P_{\mathrm e}(r)$ is the electron pressure profile, $\sigma_{T}$ is the Thomson cross section, $m_{\rm e}$ is the electron rest mass, and $g_{\nu}$ is a function of frequency\footnote{Since relativistic corrections only affect the detection significance of the most-massive clusters in the Universe at the few per cent level \citep{erler18}, for simplicity, we neglect them here and use a temperature-independent tSZ spectrum.}  written as
\BE
g_{\nu}=x\coth\left(\frac{x}{2}\right)-4,
\EE
with $x\!=\!h\nu/k_{\rm B}T_{\mathrm{CMB}}$ $\simeq$ $\nu/(56.78\,\mathrm{GHz})$ for $T_{\mathrm{CMB}}\!=\!2.725$ K. 
Equation~(\ref{EQ:TEMP-DECR}) can thus be written as
\BE
\Delta T_{\mathrm{SZ}}(\theta)=273\,\mathrm{\mu K\,}g_{\nu}\left[\frac{P_{\rm e}^{\mathrm {2D}}(\theta)}{25\,\mathrm{eV\, cm^{-3}Mpc}}\right]\;,
\EE
where $P_{\rm e}^{\mathrm{2D}}(\theta)$ is a short for the
integral appearing in the rhs of equation~(\ref{EQ:TEMP-DECR}).
Following a common practice, 
we write the pressure profile as the product of two terms: one that
describes its overall amplitude as a function of the cluster mass and redshift, and another, ${\mathcal P}$, that describes its shape:
\BEA
\nonumber
P_{\mathrm e}(r)=1.65\times10^{-3}E(z)^{8/3}\left(\frac{M_{500}}{3\times10^{14}\,h_{70}^{-1}\,\mathrm{M}_{\odot}}\right)^{2/3+\alpha_{\rm P}}\\
\times {\mathcal P}(x)\, h_{70}^{2}\,\mathrm{keV\, cm^{-3}}\;.
\label{EQ:PRESSURE-PROFILE}
\EEA
Here, $\alpha_{\rm P}$ $\simeq$ 0.12 characterizes the deviation from self-similar mass scaling (see also Section \ref{sec:mass}), $x = r/r_{\rm s}$ where $r_{\rm s}$
denotes the scale radius that defines the gas-concentration parameter
$c_{500}=R_{500}/r_{\rm s}$, and the function $E(z)\equiv H(z)/H_0$ returns the ratio between the Hubble parameter at a certain redshift and its present-day value.
The shape of the pressure profile in actual
\citep[e.g.][]{arnaud10, sun11} and simulated  \citep[e.g.][]{kay12, gupta17b} clusters is well approximated by
a generalized NFW profile \citep[GNFW,][]{nagai07},
\BE
{\mathcal P}(x)=\frac{P_{0}}{x^{\gamma}\,(1+x^{\alpha})^{(\beta-\gamma)/\alpha}}\;,
\label{EQ:GNFW}
\EE
where $P_0$ is a normalisation constant, 
while $\gamma,\:\alpha,\:\mathrm{and\:}\beta$ indicate the central ($r\!\!\ll\!\! r_{\rm s}$), intermediate ($r\!\!\thicksim\!\! r_{\rm s}$), and outer slope ($r\!\!\gg\!\! r_{\rm s}$) of the profile, respectively.
We adopt the parameters obtained in
\cite{arnaud10} 
by fitting the observed average scaled profile in the radial range $0.03<r/R_{500}<1$ combined with the average simulation profile beyond $R_{500}$:
\BE
\nonumber
\{P_{0},c_{500},\gamma,\alpha,\beta\}=\{8.40\,h_{70}^{-3/2},1.18,0.31,1.05,5.49\}.
\EE
Extensions to the GNFW profile with additional free parameters are discussed in \citet{gupta17b}.

We obtain $\triangle T_{\mathrm{SZ}}(\theta)$ as a function of $M_{500}$ and $z$ by combining equations (\ref{EQ:TEMP-DECR}) and (\ref{EQ:PRESSURE-PROFILE}).  In practice, we limit the integration
along the line of sight by truncating the pressure profile at $r_{\rm out}\!=\!6\,R_{500}$. 
We then convolve $\triangle T_{\mathrm{SZ}}(\theta)$
with the telescope beam and determine the peak temperature decrement $S(M_{500},z)$ as well as the corresponding 
signal-to-noise ratio $S(M_{500},z)/\sigma_{\rm b}$ where $\sigma_{\rm b}$ denotes the beam noise.

\begin{figure}
\centering
\includegraphics[width=8cm]{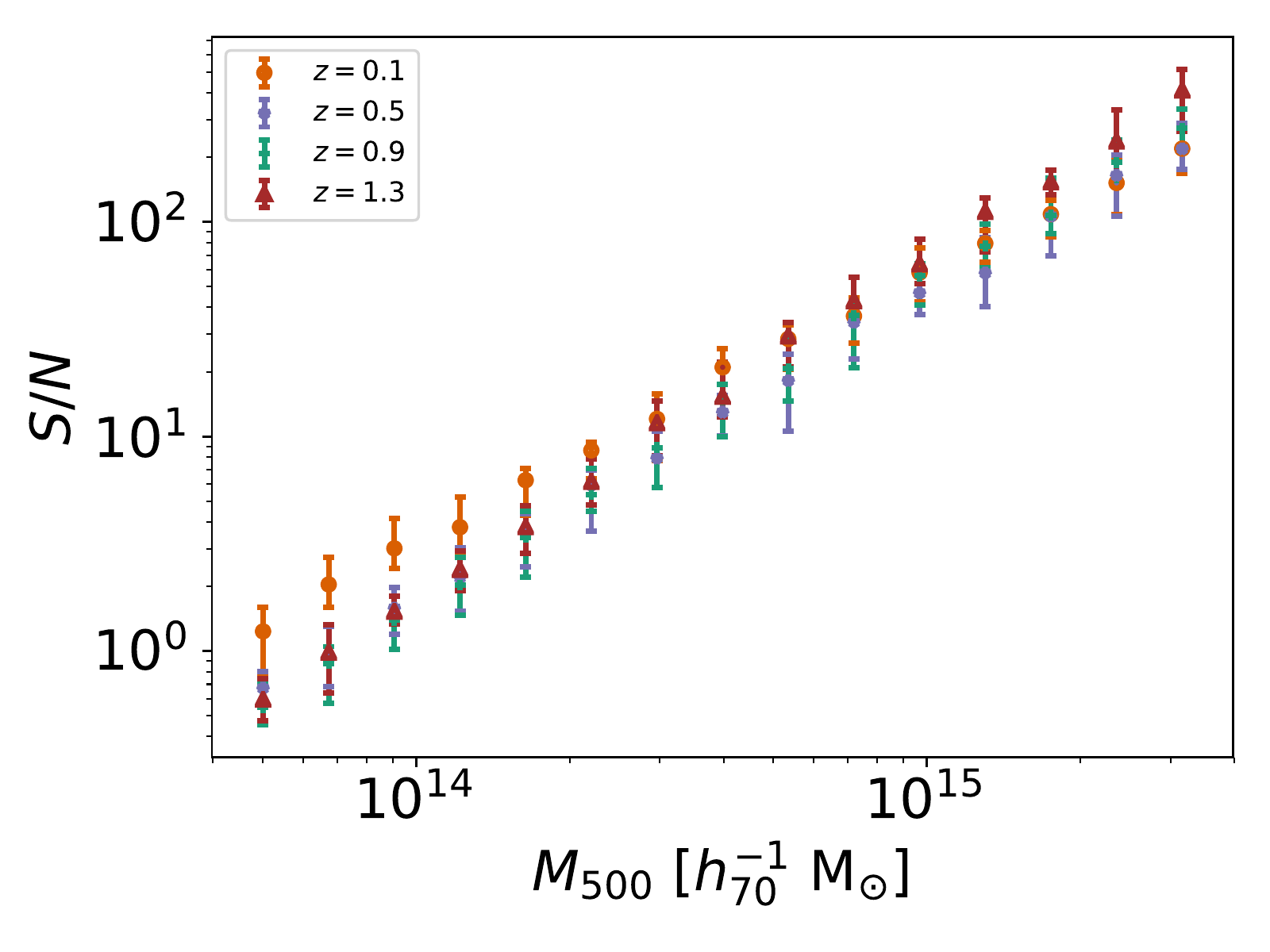}
\caption{Signal-to-noise ratio as a function of cluster mass at four different redshifts for our fiducial survey. Errorbars indicate the range between the 16$^{\rm th}$ and the 84$^{\rm th}$ percentiles, while data points show the median values.} 
\label{FIG:SN_mass}
\end{figure}

So far we have only considered the average cluster pressure profile and neglected any scatter around it. 
However, \citet{arnaud10} estimate that the
dispersion about the mean is approximately 30 per cent beyond 0.2 $R_{500}$. 
We model this by
assuming that $P_0$ is a random deviate that follows a Gaussian distribution.
Therefore, only the clusters with 
\BE
P_0>P_0^{\rm min} = P_0 \,\frac{\sigma_{\rm b}}{S(M,z)}\, S/N , 
\EE 
will be included in a catalog with a given signal-to-noise threshold $S/N$. 
It follows that the cumulative completeness can be written as an error function defined by the integral
\BE
f_{\rm compl}(M,z) = \intop_{P_0^{\rm min}}^{\infty}\frac{1}{{ \sqrt {2\pi }\,\sigma_{P_0} }}\,\exp\left[-{\frac{  \left( {x - P_0 } \right)^2 } {2\,\sigma_{P_0} ^2 }}\right]\,{\mathrm d}x,
\label{EQ:COMPLETENESS}
\EE
where $\sigma_{P_0}=0.3\,P_0$. 
With this equation, we can estimate the limiting mass for different completeness levels.
Note that the results depend on the cosmological parameters (through $D_{\mathrm A}$ and $E$) as well as on the value of $\alpha_{\rm P}$ (see Section \ref{sec:mass}).

Our calculation of the limiting mass considers only the integrated tSZ signal within the telescope beam, i.e. takes clusters in a survey as single hot pixels. There is no information on the cluster spatial profile. In actual tSZ surveys, clusters are extracted using a matched-filtering approach that takes into account both the signal shape and the tSZ spectral information, with realistic noise and foreground templates. Since our single-frequency approximation considers only thermal (white) noise, it is important to demonstrate that this approach gives realistic results. We thus apply our simplified method to the specifications of the SPT-SZ survey \citep{bleem15}. 
The data points in the left panel of Fig.~\ref{LIMITING_MASS} indicate clusters with $S/N \geq 4.5$ in the 2500 $\deg^2$ survey \citep{bleem15}
while the dot-dashed curve is the result of our calculations for
50 per cent completeness \citep[obtained using the mean values of the cosmological parameters from][and assuming $\alpha_{\rm P}=0.12$] {dehaan16}.
 
In the redshift range  $0.25<z<1.5$,
we find $\sim 700$ clusters above the mass limits computed as above,
in good agreement with the 677 clusters that have been actually detected \citep{bleem15}. 
Note that the turnover towards low limiting masses at low redshifts ($z\lesssim 0.25$) is an artifact of ignoring the atmospheric noise (also known as the $1/f$ noise) from our noise model; in reality, such low-redshift, low-mass clusters are not recovered from ground-based surveys due to their large size and lower surface brightness. 

\subsection{Fiducial survey}
\label{sec:survey}
As our fiducial survey design, we consider the planned 1000 deg$^2$ sub-mm survey for the CCAT-p telescope, following the examples presented in \citet{mittal18} and \citet{erler18}. Specifically, we use the proposed 150 GHz survey sensitivity of 6.4 $\mu$K-arcmin (corresponding to an integration time of 4,000 h) for our fiducial forecasts.
By adopting the best-fit cosmological parameters from \citet{planck16-CMB} and making use of equation (\ref{EQ:COMPLETENESS}), this sensitivity roughly corresponds to a limiting mass of $2\times 10^{14}$ M$_{\odot}$ at $z\sim 0.3$ for 50 per cent completeness (see the dashed curve in the left panel of Fig. \ref{LIMITING_MASS}).
Note that similar extremely low noise rms is already achieved with the current-generation surveys, e.g., \citet{huang19} report the depth of the new SPTpol 100 deg$^2$ map to be 6.5 $\mu$K-arcmin at 150 GHz, which is almost identical to our choice (although our survey area is ten times larger), and with completeness levels similar to those we presented in the right panel of Fig. \ref{LIMITING_MASS}. For instance, we find a limiting mass of $\sim1.7 \times 10^{14}~h^{-1}_{70}$ M$_\odot$ at $z=0.25$ which is similar to the minimum mass estimated at 50 percent completeness \citep[i.e. $\sim1.8 \times 10^{14}~h^{-1}_{70}$ M$_\odot$, see Fig. 5 in][]{huang19}.

In Fig.~\ref{FIG:SN_mass}, we derive a scaling relation for the signal-to-noise ratio with cluster mass, which is commonly used in  cosmological studies with SZ survey data \citep[e.g.][]{dehaan16, bocquet19}.
As expected, the scaling between $S/N$ and $M_{500}$ follows approximately a power law and shows a weak dependence on redshift. This further demonstrates that our simplified scheme for determining the $S/N$ and the limiting mass produces realistic results.

Although the CCAT-p survey concept has undergone several revisions and might not carry the 150 GHz option \citep[e.g.][]{choi19}, 
it is worth stressing that 
the methods and forecasts presented in this paper are not specific to any particular instrument, but rather are representative of the current and upcoming generation of tSZ surveys. 
For this reason, we consider three variations of the fiducial survey
summarised in Table \ref{TAB:DW-SURVEY-PARAMS}. The first option is to go deeper in a smaller survey area of 200 deg$^2$
keeping the total observing time unchanged, which can be suitable for telescopes with a smaller field of view and hence less mapping efficiency. Scaling from the fiducial survey depth, we get a sensitivity for this `Deep' survey of $2.8~\mu$K-arcmin. 
The second possibility is to cover a wider area of 10,000 deg$^2$ 
(roughly 24 per cent of the extragalactic sky)
with the same survey time.
This is similar to the area currently being mapped by advancedACT \citep{deBernandis16}. The sensitivity of this `Wide' survey suffers roughly 3-fold with respect to the fiducial one, reducing to $20.2~\mu$K-arcmin. The final alternative we consider is a wide-area survey with an integration time of 10,000 hours (which we call `Deep$+$Wide') that results in a sensitivity of $12.8~\mu$K-arcmin. 

We assume that redshift information will be available for all the clusters in the survey whose detection significance is above the threshold specified for our analysis.

\begin{table}
\begin{center}
\caption{Main characteristics of the four tSZ surveys considered in our forecasts. From left to right: name, integration time, survey area, beam noise and number of expected clusters with $z\leq 2$.} 
\begin{tabular}{l c c c c}
\hline
Survey  & $T_{\rm int}$ (kh) & $A$ ($\deg^2$) & $\sigma_{\rm b}$ ($\mu$K$^\prime$) & $N$ ($z\leq 2$) \\
\hline
Fiducial &4 & 1,000  & 6.4 & 2100 \\
Deep & 4 & 200 & 2.8 & 1980 \\
Wide & 4 & 10,000 & 20.2 & 2160 \\
Deep$+$Wide & 10 & 10,000 & 12.8 & 5600 \\
\hline
\end{tabular}
\label{TAB:DW-SURVEY-PARAMS}
\EC
\end{table}

%
\subsection{Cluster counts}
\label{sec:mass}
Theoretical predictions for the number density of galaxy clusters as a function of redshift are based on numerical simulations.
In order to compute the comoving number density of clusters per unit mass,
$\rd n/\rd M_{500}$, we use the fitting function for the halo mass function derived by \citet{tinker08}.
As a function of the integrated tSZ signal, $Y_{500}$,
the cluster counts
can be schematically written as 
\BEA
\frac{\rd^2 N}{ \rd Y_{500} \,\rd z} =  \frac{\rd V}{\rd z}\,\int
\frac{\rd n}{\rd M_{500}}\,
{\mathscr P}\left(Y_{500}| M_{500},z\right)\,\rd M_{500}\;,
\label{EQ:MF2OF}
\EEA
where $V(z)$ denotes the surveyed comoving volume within redshift $z$ and the function
${\mathscr{P}}\left(Y_{500}| M_{500},z\right)$ gives the conditional probability that a cluster
of mass $M_{500}$ at redshift $z$ gives a specific value of $Y_{500}$ \citep{planck13-20, planck16-24}.
In particular, we assume that $\ln Y_{500}$ is normally distributed
with scatter $\sigma_{\ln \rm Y}$
(which we take to be independent of $z$) and mean $\ln \bar{Y}_{500}$
such that
\BEA
E^{-\beta_{\rm Y}}(z)\,\frac{D_{\rm A}^{2}(z)\bar{Y}_{500}}{10^{-4}\,\mathrm{Mpc^{2}}}=Y_{*}\,h_{70}^{-2+\alpha_{\mathrm Y}}\left[\frac{(1-b_{\rm HSE})\,\mathrm{\mathit{M}}_{500}}{6\times10^{14}\mathrm{M}_{\odot}}\right]^{\alpha_{\mathrm Y}}.
\label{EQ:SCAL-RELATION}
\EEA
This scaling relation depends on the
background cosmology in terms of the angular-diameter distance
$D_{\rm A}(z)$ and $E(z)$.
Furthermore, $b_{\rm HSE}$ denotes
the bias of cluster-mass estimates
based on the assumption of hydrostatic equilibrium.

\begin{figure}
\centering
\includegraphics[width=9cm]{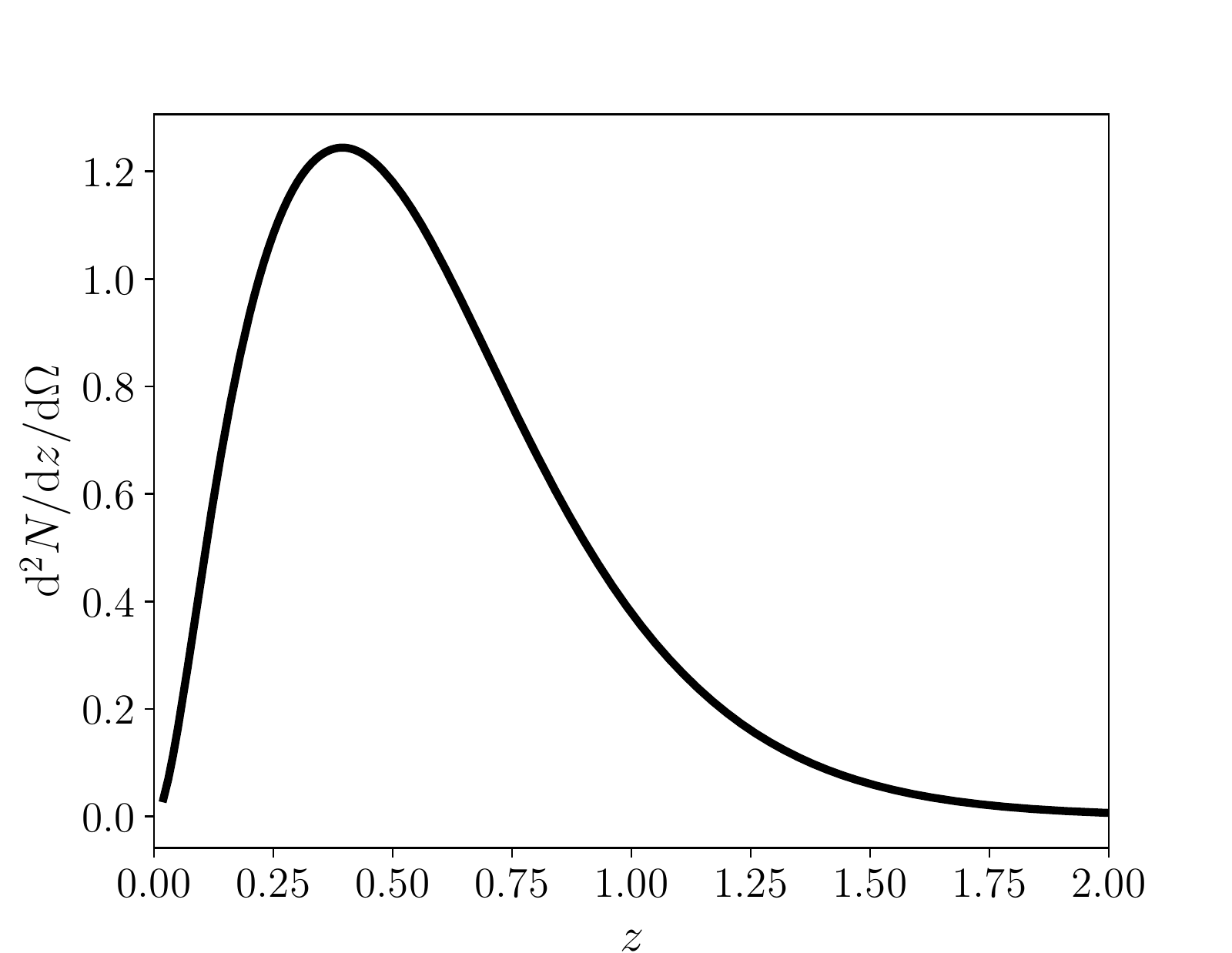}
\caption{Redshift distribution of the clusters with $S/N\geq 4.5$ in our fiducial survey.}
\label{FIG:NUMBER-COUNT}
\end{figure}

The fiducial values we adopt for the cosmological parameters 
and for the cluster scaling relation are listed in
Table~\ref{TAB:PARAM}.
For consistency, we impose that
$2/3+\alpha_{\rm p} = \alpha_{\rm Y}-1$ as both $\alpha_{\rm p}$ and
$\alpha_{\rm Y}$ describe deviations from the self-similar mass scaling
of galaxy clusters (similarly, $\beta_{\rm Y}$ and the exponent of $E(z)$ in equation~(\ref{EQ:PRESSURE-PROFILE}) express the scaling with $z$).
The resulting number of clusters per unit redshift and solid angle
for our fiducial tSZ survey is shown in Fig.~\ref{FIG:NUMBER-COUNT}. 
The total number of clusters with $S/N>4.5$ detected in the different surveys presented above are reported in Table~\ref{TAB:DW-SURVEY-PARAMS}.
\begin{table*}
\footnotesize
\begin{center}
\begin{minipage}{150mm}
\caption{\label{TAB:PARAM} Fiducial values for the parameters
that define the cosmological model \citep[from][Table 4: TT,TE,EE+lowP+lensing]{planck16-CMB}, $H_0$ \citep[from][]{riess11} and the tSZ scaling relation \citep[from][]{planck16-24}. In order to perform our forecasts, we use flat uninformative priors for all parameters that are well constrained by cluster studies (i.e. $\OM$, $\SIGMA8$ and $w$).
We impose Gaussian priors on most of the other parameters based on different probes (see Section~\ref{sec:priors} and Table~\ref{TAB:PRIORS} for details). The symbol $G(\sigma)$ indicates a Gaussian prior with standard deviation $\sigma$.
Finally, we keep $\beta_{\rm Y}$ and $b_{\rm HSE}$ fixed at their currently preferred value.
As a reference, we report the current uncertainties on $\OM$, $\SIGMA8$ and $w$ as determined in the cluster-cosmology analysis by \citet{dehaan16}
and those on the tSZ scaling relation as given in \citet{planck16-24}.}
\begin{center}
\begin{tabular}{c l c c c}
\hline
Cosmological& Description & Fiducial& Prior & Current \\
parameter & &value & &error \\
\hline
$\OM$ & Total Matter Fraction	& 0.312 & Flat & $\pm$0.046 \\
$\SIGMA8$ & Normalization of $P(k)$	& 0.815 & Flat & $\pm$0.075 \\
$ w$ & Equation-of-State Parameter for Dark Energy	& -1	& Flat & $\pm$0.31	\\
$ \Omega_{\rm b}$ & Baryon Fraction	& 0.0488& Table~\ref{TAB:PRIORS}	& -- \\   
$H_0$ 	& Hubble Constant in km s$^{-1}$ Mpc$^{-1}$ & 73.8 &Table~\ref{TAB:PRIORS} & -- \\  
$n_s$ & Scalar spectral index & 0.965 & Table~\ref{TAB:PRIORS}	& -- \\
\hline
\hline
tSZ scaling & Description & Fiducial & Prior & Current\\
parameter & & value& & error\\
\hline
$\alpha_{\rm Y}$ & YM relation: Slope & 1.79 & $G(0.08)$&$\pm$ 0.08  \\
$\log Y_{*}$ & YM relation: Normalization & -0.19 &$G(0.02)$ &$\pm$0.02 \\
$\sigma_{\ln \rm Y}$ & YM relation: Logarithmic Scatter & 0.127 & $G(0.023)$& $\pm$0.023  \\
$\beta_{\rm Y} $& YM relation: Slope of $E(z)$ & 0.66 & Point & $\pm$0.50  \\
$\rm 1-b_{\rm HSE}$ & YM relation: Mean mass bias & 0.8 & Point &$_{-0.1}^{+0.2}$ \\
\hline
\end{tabular}
\end{center}
\end{minipage}
\end{center}
\end{table*}

\begin{figure}
\centering
\includegraphics[width=9cm]{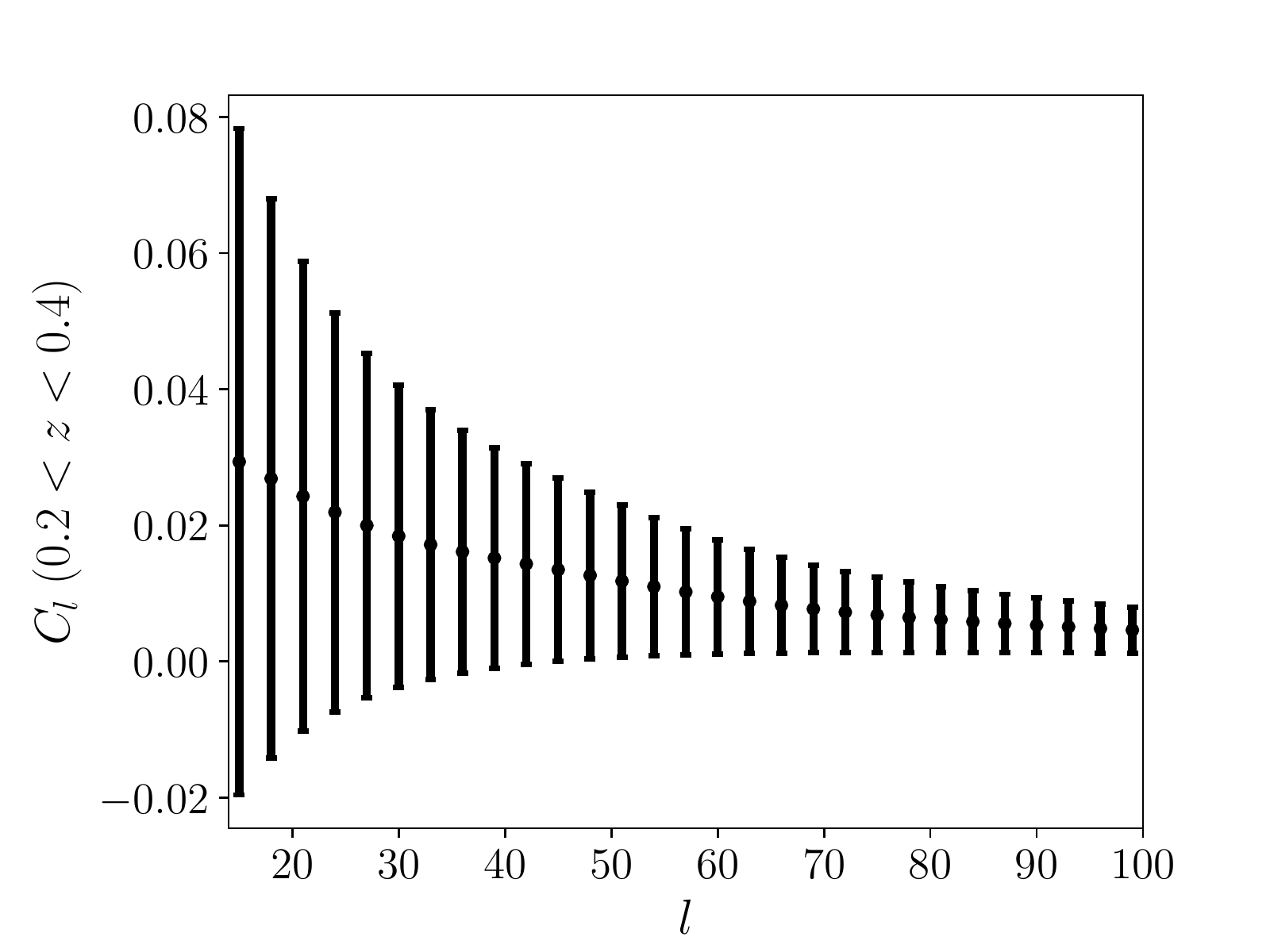}
\caption{The angular power spectrum of the clusters with $S/N\geq 4.5$
and $0.2<z<0.4$ in our fiducial tSZ survey, together with its statistical uncertainty. To improve readability, the power spectrum is averaged over multipole bins of size $\Delta l=3$. Note, however, that we consider individual multipoles to produce our forecasts.}
\label{FIG:CLUSTERING}
\end{figure}

\subsection{Angular clustering}
The second observable we consider is the angular power spectrum 
of the clusters in a number of disjoint redshift bins. 
Let $N_i$ denote the number of clusters in the $i^{\rm th}$ bin.
Using the Limber approximation, we write their power spectrum as
\BEA
C_{l}^{(i)}\simeq 4 \pi  \int
\frac{\mathrm{d}V}{\mathrm{d}z}(z)\, P\left[\frac{l+0.5}{D_{\mathrm A}(z)},\, z\right]\, W_{i}^2(z)\, {\mathrm d}z\;,
\label{EQ:LIMBER-APPROX}
\EEA
where $P(k,z)$ denotes the linear matter power spectrum and $W_i$ is
a redshift-dependent weight function,
\BE
W_{i}(z)=\frac{1}{N_{i}}\frac{\mathrm{d}N_{i}}{\mathrm{d}V}(z)\,  b_{i}\left(z\right)\;,
\label{EQ:LIMBER-WEIGHT}
\EE
that depends on the effective linear bias parameter of the clusters,
\BE
b_{i}(z)=\frac{\displaystyle{\int b_{i}(M_{500},\, z)\,\frac{{\mathrm d}^{2}N_{i}}{{\mathrm d}M_{500}\, {\mathrm d}z}(M_{500},z)\,{\mathrm d}M_{500}}}{\displaystyle{\int \frac{\mathrm{d}^{2}N_{i}}{\mathrm{d}M_{500}\, \mathrm{d}z}(M_{500},z)\,{\mathrm d}M_{500}}}\;,
\label{EQ:BIAS-Z}
\EE
where the bias coefficients for the individual haloes, $b_{i}(M_{500},\, z)$, are computed following \cite{tinker10}.

As an example, in Fig.~\ref{FIG:CLUSTERING} we show the resulting power spectrum in the redshift bin $0.2\leq z<0.4$ for our fiducial survey. 
Errorbars denote statistical uncertainties (see Section~\ref{sec:likac} for details).
Note that the signal-to-noise ratio of the measurements is rather low at all scales.
 
\subsection{Statistical analysis}
\label{sec:likelihood}
In order to forecast parameter constraints, we proceed as follows.
We first generate mock data
for our observables 
by assuming the parameter set reported in Table~\ref{TAB:PARAM}.
We then use Bayesian statistics to
fit models-- i.e. equations~(\ref{EQ:MF2OF}) and (\ref{EQ:LIMBER-APPROX})-- to the artificial data and derive the posterior distribution function for the cosmological and scaling-relation parameters. 
The last step is performed using the MCMC code {\tt emcee} \citep[a Python implementation of an affine invariant ensemble sampler;][]{mackey13}. 
Joint marginalised constraints on parameter pairs are obtained using the {\tt pygtc} code \citep{bocquet16triangle}. 

In parallel, we also derive parameter constraints using the Fisher information matrix \citep[see e.g.][]{pillepich12}. 
A comparison of the results derived with the two methods is presented in Section~\ref{sec:Fish}.

\subsubsection{Likelihood function for the number counts}
\label{NCs}
We count tSZ clusters using
20 equal bins in $\log Y_{\mathrm{500}}$ and 40 equal bins in redshift covering the range $0.01<z<2$. 
Assuming that number counts follow the Poisson distribution, we write the likelihood function for the model parameters as\footnote{This expression neglects correlations between different redshift bins due to large-scale clustering. The approximation is suitable for galaxy-cluster studies \citep{hu03a}.}
\BE
{\mathcal L} = \sum_{i} 
N_{i}^{\rm d} \ln(N_{i}^{\rm m}) - N_{i}^{\rm m} - \ln(N_{i}^{\rm d}!)\;,
\label{EQ:LIKNC}
\EE
where the sum runs over the 800 bins while $N_{i}^{\rm m}$ and $N_{i}^{\rm d}$ denote the model prediction and the actual value of the number of clusters found in the  $i^{\rm th}$ bin, respectively. Note that, for a given set of mock data, the last term on the rhs of equation~(\ref{EQ:LIKNC}) adds a constant to the likelihood function and can be neglected.

\subsubsection{Likelihood function for the angular clustering}
\label{sec:likac}
We consider the angular power spectrum of the clusters
in eight consecutive redshift bins with boundaries 
$(0.01, 0.2, 0.4, 0.55, 0.65, 0.75, 0.9, 1.1, 2.0)$.
In order to reduce shot noise, we do not further split the cluster sample based on $Y_{500}$.
We only take into account multipoles in the range $15\leq l\leq 95$. The lower limit is set since our fiducial survey covers an area of 1,000 $\deg^2$. The upper limit instead reflects the scale at which non-linearities in the matter power spectrum and in the halo bias as well as halo exclusion effects make the modelling more problematic \citep[e.g.][]{pillepich10,pillepich12}.

Assuming that the power-spectrum estimator has Gaussian errors,
we write the likelihood function as
\BE
-2\ln {\mathcal L}=\sum_i\,\sum_{l,l'}\Delta C_{l}^{(i)}\,\mathrm{Cov}^{-1}(\widetilde{C}_{l}^{(i)},\widetilde{C}_{l'}^{(i)})\,\Delta{C_{l'}^{(i)}}\;,
\label{EQ:CHI2-ANG-CLUST}
\EE
where the first sum runs over the redshift bins and the second over the multipoles. In the expression above, $\Delta C_l^{(i)}$ denotes the difference between the model predictions and the mock data.
Since we focus on large angular separations, for simplicity,
we only consider the Gaussian contribution to the covariance matrix, i.e.
\BE
{\rm Cov}(\widetilde{C}_l,\widetilde{C}_{l'})=\frac{1}{f_{\rm sky}}\,\frac{2 \widetilde{C}_l^2}{2l+1}\,\delta_{ll'}\;,
\EE
where $f_{\rm sky}$ is the fraction of the sky covered by the survey,
$\widetilde{C}_{l}^{(i)}=C_l^{(i)}+n_i^{-1}$ 
(with $n_i$ the average number of clusters per steradian)
denotes the auto spectra including a shot noise contribution (assumed to be Poissonian) and $\delta_{ll'}$ is the Kronecker symbol.

\begin{table}
\begin{center}
\caption{Gaussian priors used in our study for the cosmological parameters. 
}
\begin{tabular}{ccc}
\hline
Parameter  & Rms size & Reference	\\
\hline
$\Omega_{\mathrm{b}}h^{2}$ & 0.002 &  \cite{steigman08}\\
$H_0$ & 2.4 & \cite{riess11}\\
$n_s$ &0.006 & \cite{planck16-CMB}\\
\hline
\end{tabular}
\label{TAB:PRIORS}
\EC
\end{table}

\subsubsection{Priors}
\label{sec:priors}
Several cosmological parameters ($\OM$, $\SIGMA8$ and $w$) are well constrained by our mock data. For these variables, we use uninformative flat priors.
However, we impose narrow Gaussian priors on $\Omega_{\mathrm b}, H_0$ and
$n_s$ that are not constrained by cluster counts and/or clustering studies.
Namely, for $\Omega_{\rm b}h^2$, 
we use a prior based on the abundance of primordial deuterium combined with the standard model of big-bang nucleosynthesis \citep{steigman08};
for $H_0$, a measurement calibrated on the study of 600 Cepheid variables with the  Hubble Space Telescope; finally, for $n_s$, we use the results from the analysis of CMB temperature anisotropies in \cite{planck16-CMB}.
A summary of these priors is given in Table~\ref{TAB:PARAM}.

We also use Gaussian priors--with a size that matches the current error bars as given in Table \ref{TAB:PARAM}--for some of the parameters that define the scaling relation between $Y_{500}$ and $M_{500}$. 
A note is in order here. Our simplified analysis assumes that $\beta_{\rm Y}$ and $b_{\rm HSE}$ are known exactly while the last column in Table~\ref{TAB:PARAM} shows that their current estimates are rather uncertain. In practice, our choice is equivalent to imposing
a point (Dirac-delta) prior on them. 
Although this violates the so-called Cromwell's rule\footnote{Nothing should be assigned zero prior probability unless it is logically impossible \citep{Lindley91}.} 
in statistics, our approach should be intended as a practical way to prevent degeneracies between sets of parameters that cannot be constrained individually (e.g. $b_{\rm HSE}$ and
$Y_*$ are fully degenerate if one just uses information related with the tSZ scaling relation). 
Basically, we are using a simpler theoretical model with reduced freedom.
Our focus is on the cosmological parameters
and the main conclusions of our work are not affected by these simplifications.

\section{Results}
\label{sec:results}
In this section, we present the main results of our study.

\subsection{Constraints from the fiducial survey}
\label{sec:errors}
\begin{figure}
\centering
\includegraphics[width=9cm]{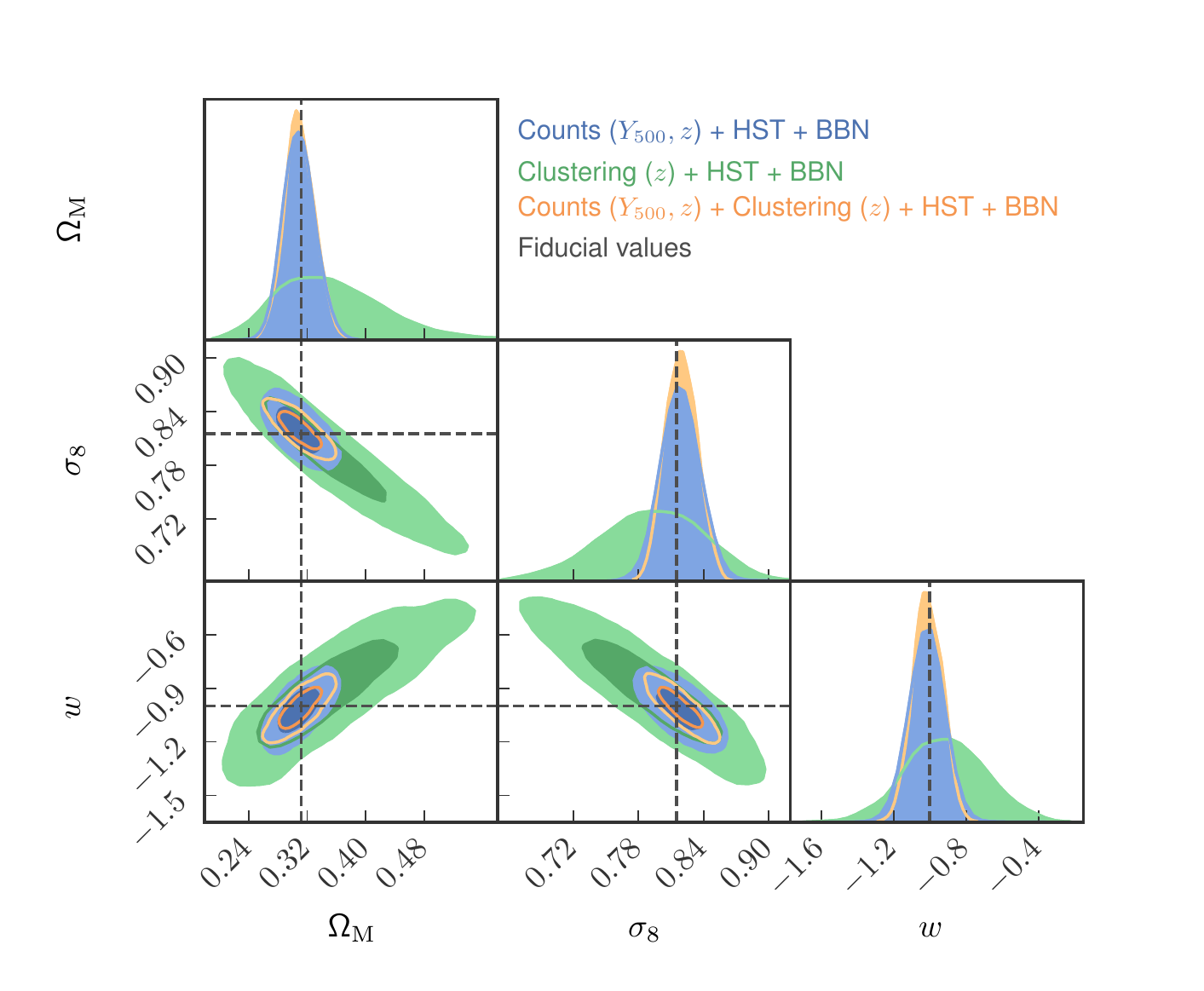}
\caption{Marginalized posterior probabilities for the
cosmological parameters derived from the fiducial tSZ survey.
Shown are the 68 and 95 per cent credibility intervals for the
joint posteriors of two parameters and the full posterior probability density for individual parameters. 
We consider three probes: cluster number counts (blue), angular clustering (green), and the combination of the two (orange). The dashed black lines represent the fiducial values of the parameters given in Table \ref{TAB:PARAM}.}
\label{FIG:CONSTRAIN-wCDM}
\end{figure}

We first discuss the constraints on the cosmological parameters set by the fiducial tSZ survey with 6.4 $\mu$K-arcmin noise and 1000 deg$^2$ area.
In Fig.~\ref{FIG:CONSTRAIN-wCDM}, we show the marginalized posterior probability distributions from the MCMC analysis. The corresponding
68 per cent credibility intervals for the individual parameters are listed in the top part of Table~\ref{TAB:CONSTRAINTS-wCDM}.
The cluster number counts constrain $\OM$, $\SIGMA8$ and $w$ at the 6.7, 2 and 8 per cent level, respectively. For each of them, this leads to a factor of a few improvement relative to current cluster samples \citep[e.g.][]{dehaan16}.
Note that, by fixing $w$ to its fiducial value and thus assuming that the accelerated expansion of the Universe is driven by 
the cosmological constant, improves the constraints on the fairly degenerate parameters $\OM$ and $\sigma_8$ to $\Delta \OM=0.015$ and $\Delta \sigma_8=0.015$. 

The angular power spectrum of the clusters sets looser constraints on the parameters (by approximately a factor of three) with respect to the number counts. This is mainly due to the large statistical uncertainties of the $C_l$'s (see Fig.~\ref{FIG:CLUSTERING}) that reflect both
the relatively small area covered by our fiducial survey ($f_{\rm  sky}\simeq 0.024$) and the
large shot-noise contributions caused by the fairly low number of clusters (a few hundreds) in each redshift bin.  

Combining the number counts with the angular clustering does not
break the degeneracies between the parameters and thus does not
lead to any significant improvement in the cosmological constraints
with respect to considering the number counts only.
We also find that the combined data cannot set competitive constraints on the parameters that define 
a time-varying equation of state for dark energy \citep[e.g. as in][]{chevallier01, linder03}. 

\begin{table*}
\begin{center}
\caption{Central 68 per cent credibility intervals for the cosmological parameters derived from the different tSZ surveys
introduced in Table~\ref{TAB:DW-SURVEY-PARAMS}.
The quoted errors refer to the distance between the median value and either the $16^{\rm th}$ or the $84^{\rm th}$ percentile of the marginalised posteriors. The errors in the central section of the table are computed under the assumption of Gaussian posteriors as described in Section~\ref{sec:errors}.}
\label{TAB:CONSTRAINTS-wCDM}
\BC
\begin{tabular}{c c c c c c c}
\hline
tSZ survey &
Probes & Additional probes& $\Delta\OM$ &  $\Delta\SIGMA8$  & $\Delta w$  \\
\hline
\\ [-0.25cm]
Fiducial & Counts($Y_{500},z$) &- 
		& 0.021 & 0.017 & 0.08 \\ [0.1cm]
Fiducial & Clustering($z$) & -
		& $^{+0.078}_{-0.063}$ & $^{+0.045}_{-0.049}$ & 0.21\\ [0.1cm]
Fiducial & Counts($Y_{500},z$) + Clustering($z$) & -
		& 0.021& 0.016 & 0.08 \\ [0.1cm]
		\hline
\\ [-0.25cm]
Fiducial & Counts($Y_{500},z$) + Clustering($z$) & {\it Planck} CMB + others
		& 0.008& 0.009 & 0.03 \\ [0.1cm]
- & - & {\it Planck} CMB + others
	& 0.009& 0.017 & 0.04 \\ [0.1cm]
\hline 
\\ [-0.25cm]
Deep & Counts($Y_{500},z$) + Clustering($z$)&-
		& $^{+0.026}_{-0.024}$ & $^{+0.022}_{-0.020}$ & 0.10\\ [0.15cm]
Wide & Counts($Y_{500},z$) + Clustering($z$)&-
		& $^{+0.020}_{-0.019}$ & $^{+0.016}_{-0.015}$ & 0.08\\ [0.15cm]
Deep+Wide & Counts($Y_{500},z$) + Clustering($z$)&-
		& $^{+0.017}_{-0.015}$ & $^{+0.014}_{-0.012}$ & 0.07\\ [0.1cm]
\hline
\end{tabular}
\EC
\end{center}
\end{table*}

Finally, we extend our forecasts to include information from the analysis of CMB anisotropies and other probes of the large-scale structure of the Universe.
In this case, we use a simplified approach by assuming
that the posterior distribution is approximately Gaussian close to its peak.
We thus download the
`base-w plikHM+TTTEEE+lowTEB+BAO+H070p6+JLA' chain from the {\it Planck} web site and compute the covariance matrix $\boldsymbol{\mathsf {C}}_{\rm ext}$ of the cosmological parameters considered in our investigation.
Similarly, we compute the covariance matrix $\boldsymbol{\mathsf {C}}_{\rm tSZ}$ resulting from the number counts and the clustering measurements with our fiducial
tSZ survey.
We estimate the parameter covariance resulting from the combined data as $\boldsymbol{\mathsf {C}}_{\rm tot}=(\boldsymbol{\mathsf {C}}^{-1}_{\rm ext}+\boldsymbol{\mathsf {C}}^{-1}_{\rm tSZ})^{-1}$.
Following this procedure, we find that the full dataset
constrains $\Omega_{\rm m}$, $\sigma_8$ and $w$ at the 2.5, 1.1 and 3 per cent level, respectively.
Note that the marginal error on $\Omega_{\rm m}$ is mostly determined by the external data, whereas those on $\sigma_8$ and $w$ are significantly improved by the tSZ experiment. 

%
\begin{figure}
\centering
\includegraphics[width=9cm]{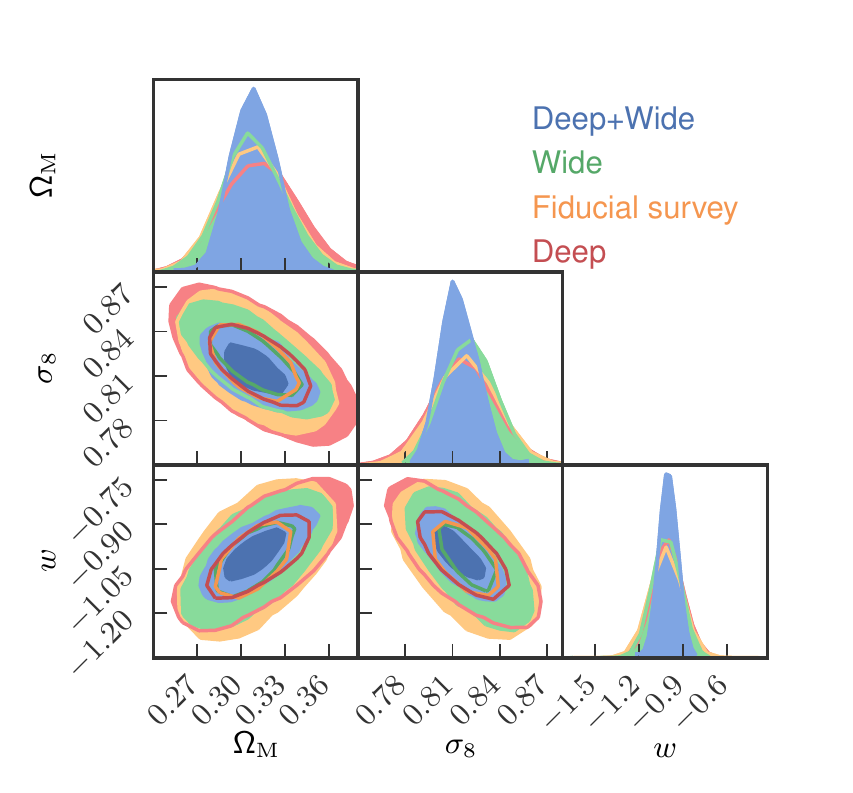}
\caption{As in Fig.~\ref{FIG:CONSTRAIN-wCDM} but for the different survey strategies described in Table~\ref{TAB:DW-SURVEY-PARAMS}. In all cases, we show parameter constraints obtained by combining the data on cluster number counts and angular clustering. 
}
\label{FIG:wide_deep}
\end{figure}
\subsection{Fiducial vs deeper/wider surveys}
\label{sec:areas}
At fixed observing time, is it more advantageous to realize a deeper or a wider tSZ survey? We address this question by comparing
the forecasts we obtain for the different survey strategies described
in Table~\ref{TAB:DW-SURVEY-PARAMS}.
In analogy with our previous discussion,
we consider both cluster counts and angular clustering\footnote{Although, even for the `Wide' survey, accounting also for angular clustering only leads to very minor improvements.}
and we calculate the selection functions for the different telescope sensitivities using the technique discussed in Section~\ref{sec:limmass}. Our results are presented in
Fig.~\ref{FIG:wide_deep} and Table~\ref{TAB:CONSTRAINTS-wCDM}. 
We find that the `Wide' survey places slightly tighter 68 per cent credible intervals for the cosmological parameters than the `Deep' one.
Specifically, the `Wide' (`Deep') survey sets constraints on
$\OM$, $\SIGMA8$ and $w$ at the 
6 (8), 1.8 (2.6), and 8 (10) per cent level, respectively.
The improvement with respect to the fiducial survey is marginal.

Finally, we consider the `Deep+Wide' survey that requires 2.5 times longer integration time than the fiducial case (with the same experimental
apparatus).
As expected, this set-up puts tighter constraints on the cosmological parameters
with respect to the surveys considered above
(see Fig.~\ref{FIG:wide_deep} and Table~\ref{TAB:CONSTRAINTS-wCDM}). 

\subsection{Cosmology-dependent selection function vs fixed limiting mass}
\label{sec:varmass}
\begin{figure}
\centering
\includegraphics[width=8.5cm]{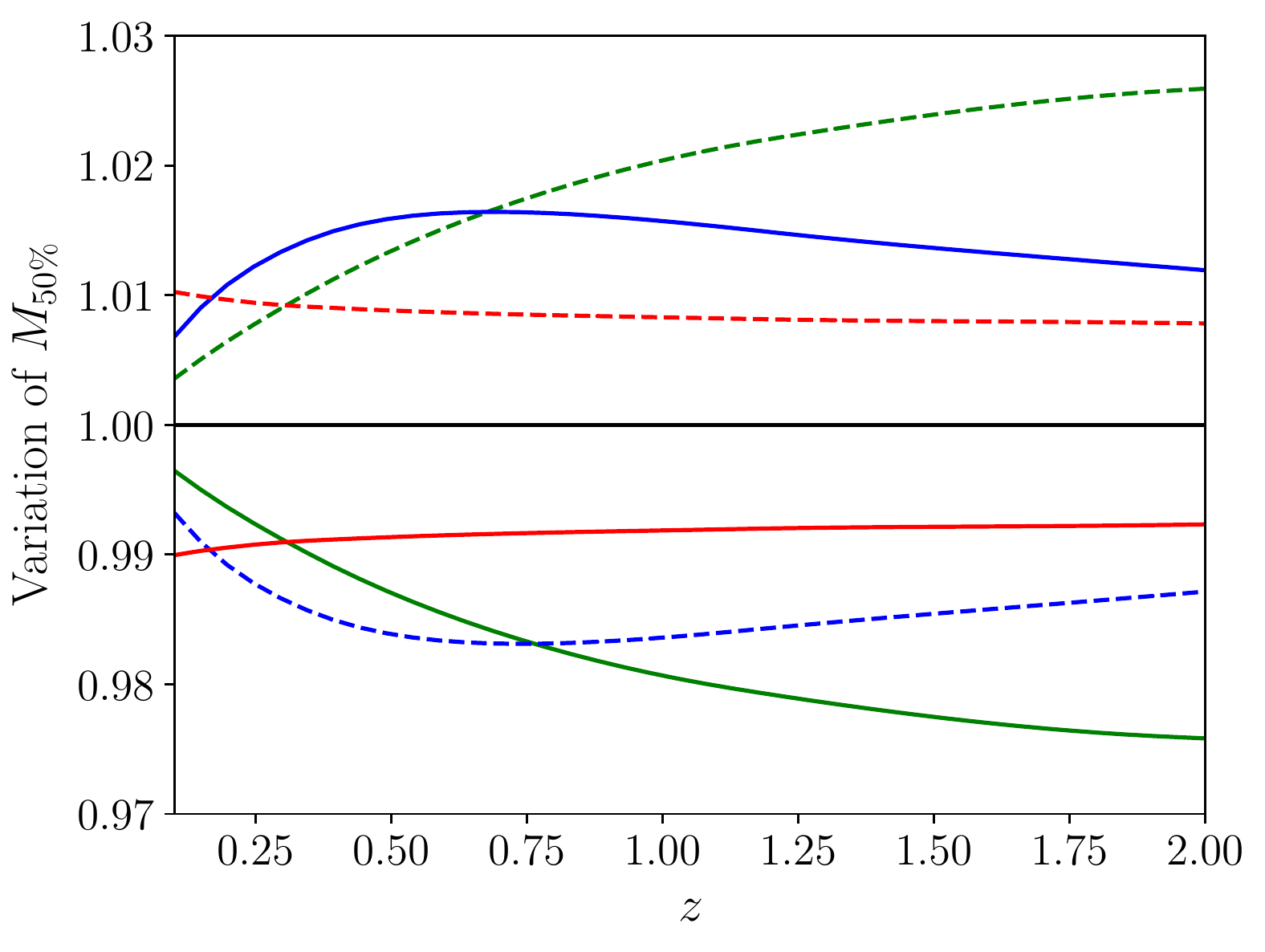}
\includegraphics[width=8.5cm]{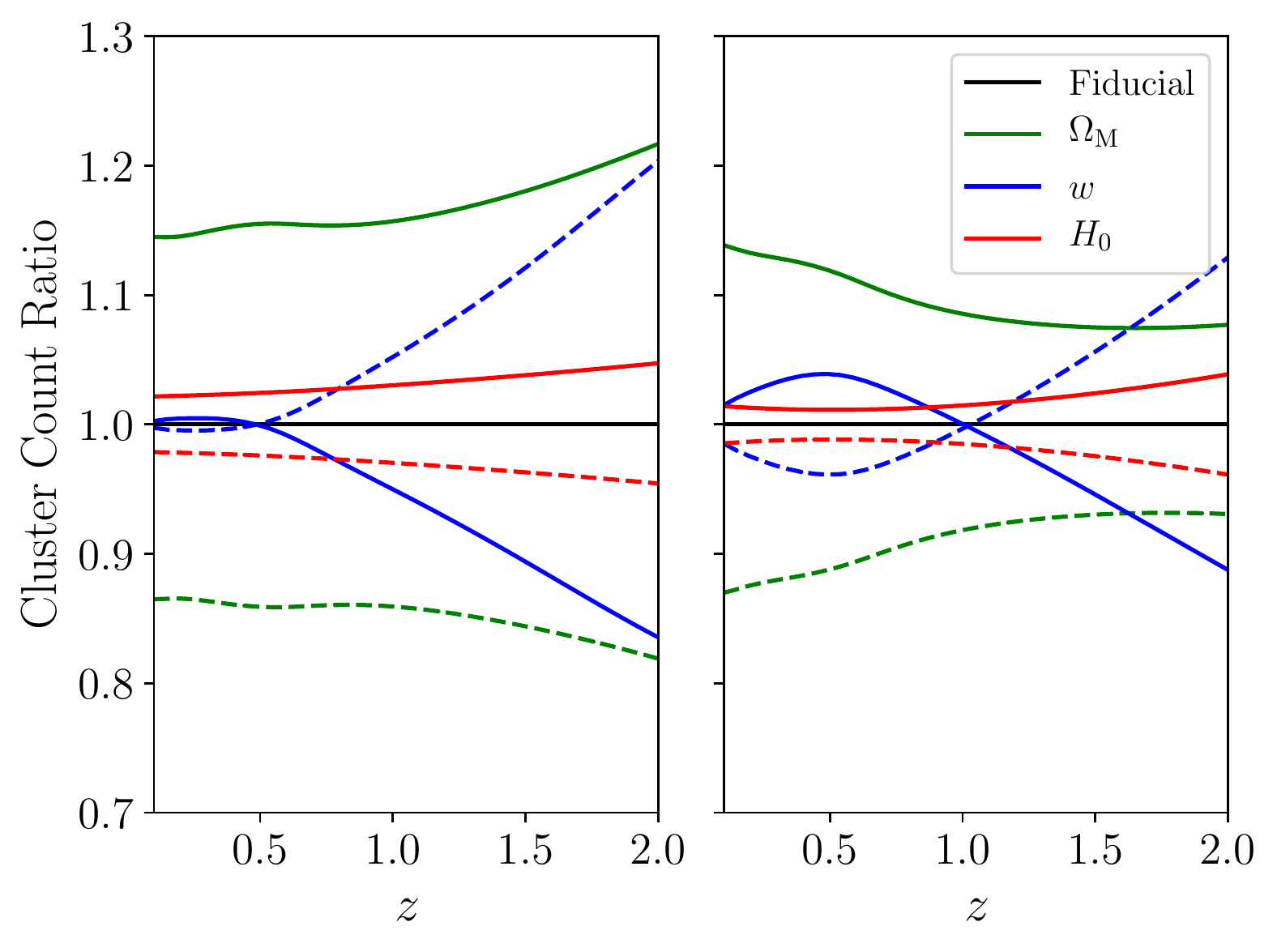}
\caption{Sensitivity of the cluster number counts measured in our fiducial survey to small variations of selected cosmological parameters.
The top panel considers 
the cluster mass that corresponds to 50 per cent completeness, $M_{50\%}$. 
Shown is 
the ratio between $M_{50\%}$ obtained assuming a modified model Universe and
that obtained within the fiducial cosmological model described in Table~\ref{TAB:PRIORS}.
We vary one parameter at a time  by $+5$ (solid) and $-5$ (dashed) per cent with respect to the fiducial value.
The bottom panels show similar ratios between the cluster counts. 
However,
in the left panel, the selection function of the tSZ survey is computed as
in Section~\ref{sec:limmass} using the modified cosmology while the mass selection for the fiducial cosmological model is always used in the right panel.
}
\label{FIG:ratio_fiducial}
\end{figure}

\begin{figure*}
\centering
\includegraphics[trim={0.2cm 0.9cm 0.2cm 0cm},clip,width=18cm, scale=0.5]{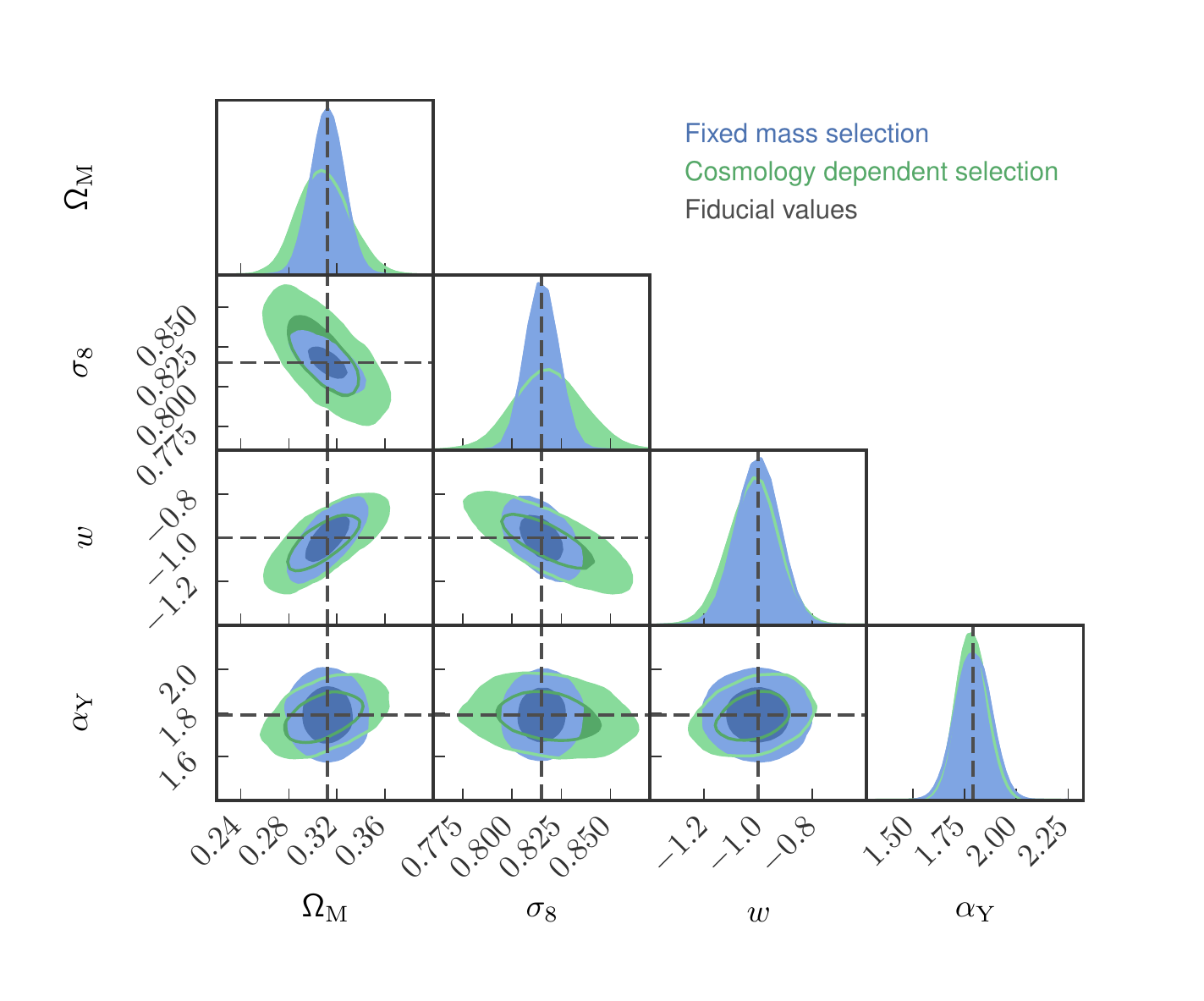}
\caption{Comparison 
between forecasts obtained with cosmology dependent (green) and fixed mass (blue) selections for our fiducial tSZ survey.
The plot is structured as in Fig.~\ref{FIG:CONSTRAIN-wCDM} but also includes $\alpha_{\rm Y}$ and only accounts for information from the cluster number counts.
}
\label{FIG:fixed_varyingLM}
\end{figure*}
Since the pioneering work by \citet{haiman01}, 
forecast studies for tSZ experiments usually assume a fixed limiting mass that does not depend on the cosmological parameters \citep[e.g.][]{holder01b, mak11, pierpaoli12, khedekar13}.
On the contrary, our analytical approximation to the survey selection function 
presented in Section~\ref{sec:limmass} makes it possible for us
to efficiently derive model constraints that account for variations in the cluster selection criteria.
We are thus in the position of using our forecasts
to investigate the impact of a cosmology-dependent selection function.

In Fig.~\ref{FIG:ratio_fiducial}, we study the sensitivity of the cluster number counts to small variations in the cosmological parameters that influence the
tSZ selection function (i.e. $\Omega_\mathrm{m}, H_0$ and $w$).
In the top panel, we consider
the cluster mass that corresponds to 50 per cent completeness, $M_{50\%}$, as a function of $z$.
Shown are the fractional changes in $M_{50\%}$
with respect to the fiducial cosmology when a single parameter is varied by $\pm 5$ per cent. 
Typically, $M_{50\%}$ is altered by a few per cent but, given the steep slope of the halo mass function, this is enough to generate substantial 
changes in the measured number counts as a function of redshift (bottom left panel).
For comparison, we also show the variations in the cluster counts that one obtains by artificially keeping the selection function fixed to the fiducial cosmology (bottom right panel). In this case, variations in the cluster counts reflect changes in the halo mass function. 
At all redshifts, the most evident differences between the bottom panels appear when $\Omega_\mathrm{M}$ or $w$ are varied.
On the other hand, changing the Hubble constant has a smaller impact although $H_0$ enters the $Y_{500}$-$M_{500}$ scaling relation directly through $D_{\rm A}^2$.
These results suggest that neglecting the cosmological dependence of the tSZ selection function might produce inaccurate forecasts
for $\Omega_\mathrm{M}$, $w$ and all the parameters that are degenerate with them, like $\sigma_8$.

We directly test this conjecture by performing a forecast for the cluster number counts of our fiducial survey in which we do not vary the tSZ selection function with cosmology but keep it fixed (using  the result for the fiducial cosmology). 
The marginalized posterior probabilities for all the model parameters are shown in Fig.~\ref{FIG:fixed_varyingLM} together with their
counterparts obtained by varying the selection function.
The sizes of the 68 per cent credibility intervals are directly compared
in Table~\ref{TAB:ratio_fiducial}.
Generally, the constraints on the cosmological parameters are
substantially weaker when we vary the selection function.
In other words, 
cosmological forecasts that use a fixed selection function yield unrealistically optimistic constraints. 
The changes are particularly significant for $\sigma_8$ and $\Omega_\mathrm{M}$. Namely, by letting the selection function vary,
$\Delta \sigma_8/\sigma_8$ increases from 0.9 to 2 per cent, whereas $\Delta \Omega_\mathrm{M}/\Omega_\mathrm{M}$ grows from 4 to 6.7 per cent. 
On the other hand, the error on the slope of the scaling relation $\alpha_\mathrm{Y}$ improves marginally.

\begin{table}
\begin{center}
\caption{Comparison between forecasts obtained with variable and fixed mass selections for our fiducial tSZ survey.
Quoted are the central 68 per cent credibility intervals
for all the model parameters that are not fully constrained by the priors listed in Tables~\ref{TAB:PARAM} and \ref{TAB:PRIORS}.
The forecasts are obtained from the cluster number counts
and the corresponding posterior probabilities are shown in Fig.~\ref{FIG:fixed_varyingLM}.} 
\label{TAB:ratio_fiducial}
\BC
\begin{tabular}{c c c c c}
\hline
\hline
   & Fixed   & Variable \\
Parameter  &  mass selection & mass selection \\
\hline
$\Delta\OM$         &  0.013    &  0.021   \\
$\Delta\SIGMA8$     &  0.007    &  0.017   \\
$\Delta w$        &  0.070     &  0.080    \\
$\Delta \alpha_{\rm Y}$ & 0.081 & 0.073    \\
\hline
\end{tabular}
\EC
\end{center}
\end{table}
\begin{figure}
\centering
\includegraphics[width=9cm]{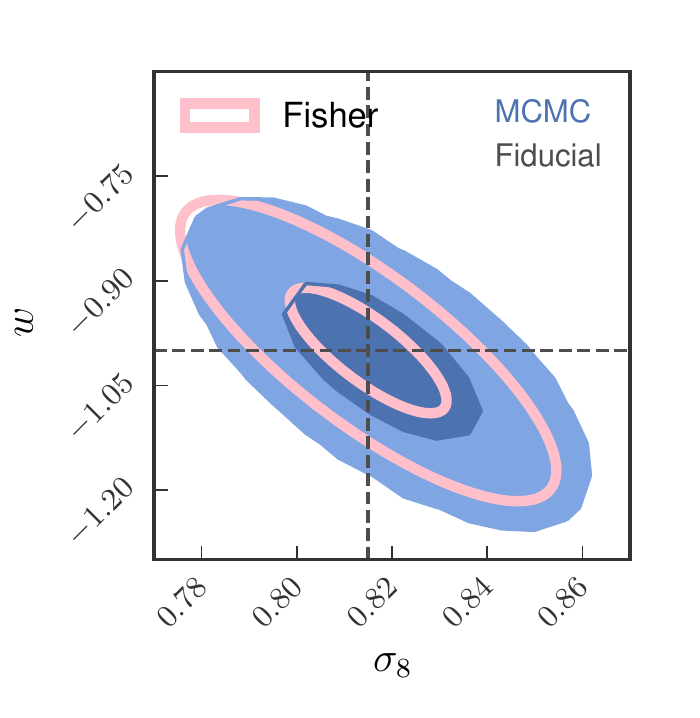}
\caption{Joint posterior distribution of $\sigma_8$ and $w$
obtained with the MCMC method (blue) and with the Fisher matrix formalism (pink). The contour levels mark the boundaries of the 68 and 95 per cent credible regions obtained from the cluster number counts in the fiducial tSZ survey.}
\label{FIG:fisher_mcmc}
\end{figure}
\subsection{MCMC vs Fisher analysis}
\label{sec:Fish}

The Fisher information matrix formalism \citep{tegmark97} provides a fast method to make forecasts for future experiments 
under the assumption that the posterior probability density of the model parameters is approximately Gaussian around its peak.
The accuracy of this method varies case by case and critically depends on the assumed experimental set-up and model parameterization \citep[e.g.][]{perotto06, wolz12, khedekar13, pillepich18}.
In order to ease comparison with previous work on tSZ experiments based on the Fisher approach \citep[e.g][]{mak11, pierpaoli12}, we recompute the forecast for the cluster number counts in the fiducial survey 
using the Fisher-matrix method. In general, we find rather good agreement with the results obtained from the MCMC simulations. For instance, the model parameters show the same degeneracies although
the marginal errors derived with the Fisher analysis are systematically smaller by approximately ten per cent.
As an illustrative example, in Fig.~\ref{FIG:fisher_mcmc} we compare the marginalised constraints we obtain in the $\sigma_8$-$w$ plane.

\section{Discussion and conclusions}
\label{sec:discussion}
This paper presents
cosmological forecasts for future `straw-man' ground-based tSZ surveys that cover a sky area between $200$ and $10^4$ deg$^2$
to an rms noise level between 2.8 and 20.2 $\mu$K-arcmin at 150 GHz. The fiducial survey parameters are based on a deep multi-frequency observational run planned for the CCAT-p telescope but we also explore wider and deeper options (see Table~\ref{TAB:DW-SURVEY-PARAMS}).
We use a redshift-dependent scaling relation to 
associate the integrated Compton-$Y$ parameter of a cluster with the mass of its host dark-matter halo and create mock data for the different survey options. We assume that the individual redshifts of the clusters are known from optical observations. 
The selection functions of the tSZ surveys are obtained with a simplified analytic method introduced in Section~\ref{sec:limmass}. This provides a fast and convenient approximation to the more accurate (but time consuming) technique that uses matched filters to extract galaxy clusters
from mock observations based on hydrodynamical simulations. 
Our simplistic approach reproduces the SPT selection function (Fig.~\ref{LIMITING_MASS}) and
generates a power-law scaling between the cluster mass and
the tSZ detection significance (Fig.~\ref{FIG:SN_mass}).

We fit theoretical models to the mock data and use MCMC simulations to derive the posterior distribution function for the cosmological parameters marginalized over the uncertainty on the $Y_{500}$-$M_{500}$ scaling relation. 
Although we use information from the cluster number counts and their angular clustering, in all cases, we find that the resulting constraints on $\OM$, $\SIGMA8$ and $w$ are dominated by the counts\footnote{
\citet{pillepich18} draw similar conclusions for the currently ongoing X-ray survey {\it eROSITA} that covers a much larger sky fraction ($f_{\rm sky}\sim0.65$) than the ground-based tSZ surveys we consider here.
Note, however, that clustering information is key to constrain the level of primordial non-Gaussianity in the matter perturbations \citep{pillepich12}.}. 
Note that we impose narrow Gaussian priors on $\Omega_{\mathrm b}, H_0$ and $n_s$ using external data sets (see Table~\ref{TAB:PRIORS}).
We find that our fiducial tSZ survey sets constraints on $\OM$, $\sigma_8$ and $w$ at the 6.7, 2 and 8 per cent level, respectively.
If the value of $w$ is fixed to $-1$, then the constraints on the first two parameters reduce to 4.8 and 1.8 per cent, respectively.  

We find that, for a fixed observing time, wider surveys provide slightly better cosmological constraints than deeper ones. 
This reflects two things: (i) a wider survey detects a larger number of massive clusters despite its poorer sensitivity and (ii) the high-mass end of the halo mass function is very sensitive to variations in the cosmological parameters.
A wider-area survey also offers more opportunities for (i)
cosmological studies based on the tSZ power spectrum and/or higher-order statistics after removing the resolved clusters \citep[e.g.][]{george15}; and (ii)
cross-correlation studies with other observational probes \citep[e.g.][]{giannantonio16}.
These benefits are not included in our analysis.
On the other hand, deeper surveys detect more 
lower-mass systems for which the mass calibration based on the $Y_{500}$-$M_{500}$ scaling relation becomes much less secure.
This issue, however, can be circumvented by using external mass calibrations based, for instance, on weak-lensing measurements \citep[e.g.][]{bocquet19}.

In this work, we use MCMC sampling to evaluate 
the joint posterior probability distributions of the model parameters.
However, in order to compare our results with previous studies based on the Fisher information matrix, we repeat some of the forecasts using this method.
For the surveys we consider here, we find that Fisher forecasts are mostly consistent with the MCMC results.
Although the cosmological parameters show the same degeneracies in the two cases, the Fisher method systematically underestimates the errors by approximately 10 percent. 
 
In cosmological forecasts based on the tSZ effect, it is customary to use a fixed signal-to-noise converted to a limiting-mass cut-off for the cluster counts. 
One of the main objectives of this paper is to show the importance of taking into account the response of the survey selection function to 
the variations of the cosmological parameters in the likelihood function.
We first demonstrate the sensitivity of the predicted cluster number counts to 
this effect and show that changes in $\OM$ and $w$ produce the largest response (Fig.~\ref{FIG:ratio_fiducial}).
It follows that, by neglecting the dependence of the survey completeness on the underlying cosmological model, one obtains artificially tighter constraints on several cosmological parameters.
In fact, after accounting for the changes in the selection function in our forecasts, 
$\Delta \sigma_8/\sigma_8$ increases from 0.9 to 2 per cent (i.e. more than doubles),
$\Delta \OM/\OM$ from 4 to 6.7 per cent, and
$\Delta w/w$ from 7 to 8 per cent.
On the other hand, the error on the slope of the scaling relation $\alpha_\mathrm{Y}$ marginally improves (see Table~\ref{TAB:ratio_fiducial} and Fig.~\ref{FIG:fixed_varyingLM}).

A final note is in order here. 
The discussion about the variations of the selection function with cosmology above is relevant not only at the level of forecasts but also in the actual data analysis of tSZ surveys. In fact, these variations  directly impact the likelihood function for the cluster number counts. While this is easy to account for with our quick and approximate analytical method, it poses a major challenge in a more realistic setting.
It would basically require using matched filters to extract clusters from mock observations for every variation of the cosmological parameters in a Markov chain. This would take a prohibitive long time. 
However, we already have strong priors on the cosmological parameters coming from previous cluster studies (as well as from the CMB and other large-scale-structure probes) and this implies
that the region of parameter space over which the variations of the selection function should be analyzed is relatively small.
Therefore, a viable approach could be to run a much smaller set of the computationally expensive matched-filter simulations to train a Bayesian emulator and then use the fast predictions of the trained model for the survey selection function together with the MCMC sampler.

\section*{Acknowledgements}

NG acknowledges support from the Australian Research Council Discovery Projects scheme (DP150103208). 

\bibliographystyle{mnras}
\bibliography{tSZ_forecasts}

\end{document}